# ON-DEVICE NEURAL SPEECH SYNTHESIS

*Sivanand Achanta\*, Albert Antony, Ladan Golipour, Jiangchuan Li, Tuomo Raitio,*
*Ramya Rasipuram, Francesco Rossi, Jennifer Shi, Jaimin Upadhyay, David Winarsky, Hepeng Zhang*

Apple

## ABSTRACT

Recent advances in text-to-speech (TTS) synthesis, such as Tacotron and WaveRNN, have made it possible to construct a fully neural network based TTS system, by coupling the two components together. Such a system is conceptually simple as it only takes grapheme or phoneme input, uses Mel-spectrogram as an intermediate feature, and directly generates speech samples. The system achieves quality equal or close to natural speech. However, the high computational cost of the system and issues with robustness have limited their usage in real-world speech synthesis applications and products. In this paper, we present key modeling improvements and optimization strategies that enable deploying these models, not only on GPU servers, but also on mobile devices. The proposed system can generate high-quality 24 kHz speech at 5x faster than real time on server and 3x faster than real time on mobile devices.

## 1. INTRODUCTION

With the recent advances in neural text-to-speech (TTS) technology, it is possible to construct a fully neural network based TTS system. For example, Tacotron [1], an attention-based sequence-to-sequence model, can generate a Mel-spectrogram from a sequence of characters or phonemes, followed by a neural vocoder, such as WaveNet [2], WaveRNN [3], or Parallel WaveNet [4], that generates speech samples conditioned on the Mel-spectrogram. The resulting system reduces the complexity of a traditional TTS system and generates high quality speech that can be indistinguishable from natural speech [1].

A fully neural network based end-to-end TTS system has many advantages in addition to higher quality. They can be trained with a large amount of text and speech data with minimum human annotation and domain-specific expertise. They do not usually require phoneme-level alignment and contain fewer components, thus preventing compounding of errors. However, it is not straightforward to construct a fully neural network-based (on-device) TTS system that is viable as a product for millions of users. There are multiple challenges and trade-offs in terms of quality, computational cost, speed, and robustness.

In this paper, we focus on a neural TTS system consisting of two neural networks. The first network is a sequence-to-sequence attention based model, similar to Tacotron [1], that predicts a Mel-spectrogram from the sequence of phonemes, word boundaries, and punctuations. The second network is an autoregressive recurrent neural network, similar to WaveRNN [3], that generates one sample at a time, conditioned on the previous sample and the Mel-spectrogram predicted by the first network. We describe the challenges faced by using Tacotron as a neural frontend alongside the design choices that allowed us to overcome them. We also explain the required optimizations and performance improvements for WaveRNN while maintaining high quality of synthetic speech.

While the latest neural network-based TTS models are usually run using high-performance cloud computing, there are many advantages to running the models on consumer devices, such as smart phones or home automation devices: better response time, off-line or low-bandwidth usage, smaller cloud resource cost, and most importantly, improved privacy. Therefore, we will go in-depth on how such computationally intensive architectures can be transformed into an on-device TTS system. Specifically, we propose a variation of WaveRNN that significantly increases the speed of generation on mobile devices.

## 2. PRIOR WORK

We briefly review various neural TTS systems that have been proposed in the recent past. Most state-of-the-art end-to-end neural TTS systems use a cascade of two separately trained networks, the first converting graphemes or phonemes into intermediate features [5, 1, 6, 7, 8, 9], and the second predicting the waveform as a neural vocoder [2, 10, 3]. Both components are often autoregressive while non-autoregressive parallel approaches are preferable in terms of computation, both in generating intermediate features [11, 12, 13, 14] and in neural vocoders [4, 15, 16, 17]. There are also systems that directly generate a waveform from a single network but use an intermediate representation [12], while some systems use fully end-to-end models without intermediate features or multiple networks [18, 14]. In this work, we focus on developing on-device TTS using two separate autoregressive networks, using Mel-spectrogram as an intermediate feature, while continuing our research on parallel architectures for future implementations.

---

\*Authors listed in alphabetical order by the last name.

## 3. MODEL ARCHITECTURE

Our model architecture is composed of two separately trained networks; Tacotron, a sequence-to-sequence network with attention [1], and WaveRNN, a single-layer recurrent neural network [3].

### 3.1. Tacotron

We use phoneme sequence in combination with punctuation and word boundaries as input to the Tacotron model. This way, we minimize the occurrence of mispronunciations while allowing the network to learn appropriate prosody and pausing through the punctuation. The output of the Tacotron model is a sequence of Mel-spectrogram frames computed through a short-time Fourier transform (STFT) using a 25 ms frame size and 10 ms frame shift. At the output of Tacotron, we predict two Mel-spectrogram frames at a time.

The original Tacotron 2 model [1] uses location-sensitive attention [1] which creates a soft alignment between the input and output sequences. In our initial experiments, we observed that the location-sensitive attention is not always robust, especially when the input sentence has repeated patterns. In TTS, the alignment between input and output sequences is always monotonic. Therefore, we experimented with *location sensitive monotonic attention*, which is a combination of location-sensitive attention and monotonic attention [19], and with *location sensitive stepwise monotonic attention* [20] that, in addition to the locality and monotonicity properties, prevents skipping over any input tokens. Our evaluations show that the stepwise monotonic attention performs the best and is not only robust to repetitive patterns in the input text but also to out-of-domain input text. Finally, we increased the size of encoder LSTMs to 512 units to further improve the output speech quality. In this paper, we call this model as Tacotron for simplicity.

### 3.2. WaveRNN

We use a modified WaveRNN [3] as our neural back-end. The original WaveRNN has been modified in two ways. First, the output signal representation has been changed from 16-bits to 8-bits with $\mu$-law companding. Second, the hidden state size has been reduced from 896 units [3] to 512 units. Additionally, for on-device implementation, we have further optimized the model using the split-state method as described in section 4.3.1.

Our motivation to find an alternative to the originally proposed 16-bit WaveRNN comes from the inference speed constraints. Our CUDA implementation of the WaveRNN model using coarse and fine bits generated speech only at 2x faster than real time. Using performance tools to assess the reason for slow inference showed that large part of the time is taken up in synchronizing various kernels in WaveRNN. For example, the fine-bit kernels of the WaveRNN have to wait until the coarse-bit has been predicted and vice-versa. This sequential dependency led us to think of an alternative which has single sampling instead of dual sampling to reduce the synchronization delays. Experiments in [21, 22] also suggest that all the 16-bits are not perceptually important, especially if an appropriate pre-emphasis filter is used to avoid excess quantization noise. Based on our experiments, WaveRNN with 8-bit $\mu$-law quantization with a pre-emphasis filter of 0.86 for 24 kHz sampling rate provides a good trade-off between the performance and quality.

With Mel-spectrogram as the conditioning input rather than the linguistic features, we discovered that the hidden state size of the gated recurrent unit (GRU) layer can be drastically reduced from 896 units to 512 units. In fact, in our experiments we found that for female speakers the WaveRNN size can be further reduced to 384 units without any loss in quality. However, for male speakers, this resulted in some quality loss, and we decided to use 512 units for the final model.

### 3.3. Bandwidth extension to 48 kHz

Using 24 kHz sampling rate provides a good performance at synthesis time but limits the frequency bandwidth to 12 kHz, and therefore the highest frequencies in the speech signal cannot be reproduced. As a result, the speech signal is perceived as having a diminished "brightness". In order to compensate for this effect after the speech waveform is synthesized, we perform a linear interpolation of the 24 kHz speech samples to get an upsampled 48 kHz signal. In effect, the 0–12 kHz spectrum is mirrored to 12–24 kHz with an appropriate slope so that the audible high frequencies in the 12–20 kHz band mostly follow the spectrum of natural 48 kHz speech signal. While the high-frequency spectrum (20-24 kHz) has unnaturally high amplitude due to mirroring, it is inaudible for humans. As a result of the above process, the bandwidth-extended audio is nearly indistinguishable from the original 48 kHz speech signal, provided that the generated 24 kHz signal is of high quality.

## 4. PERFORMANCE

In this section, we describe the neural TTS streaming architecture and performance improvements for all models to achieve low latency and high speed with minimal computation.

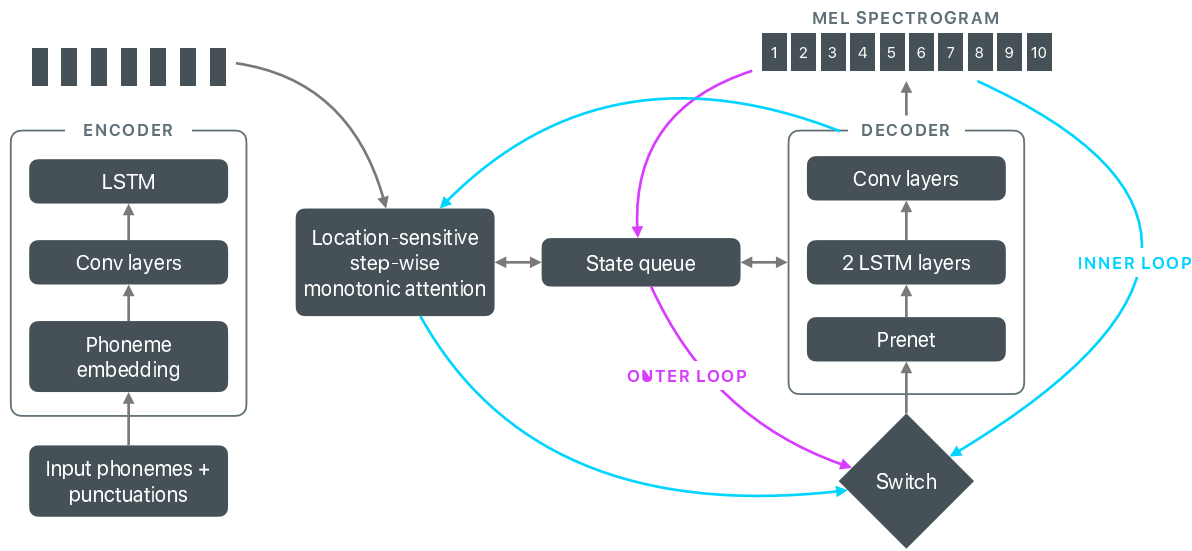

**Fig. 1**. Tacotron model with the streaming and queuing mechanisms. There are 5 executions of the inner stream loop (5 decoder steps) per one execution of the outer stream loop. For each decoding inference, we always start with outer stream loop, read the cached states first, and use it as input to the inner stream loop until the inference is complete.

### 4.1. Streaming for neural frontend and backend

The latency of the neural TTS system was originally 10 times higher than with our previous hybrid unit selection system [23]. In order to reduce the latency, we designed an end-to-end streaming approach. In the proposed streaming architecture, we perform the inference and deliver the output continuously and incrementally rather than in a serial fashion. Firstly, the encoder and decoder graphs of Tacotron are executed independently. The encoder processes the entire input, and the output of the encoder is cached. Given the encoder outputs, the decoder uses a queuing mechanism to perform the inference.

The decoder inference has two streaming loops, an inner stream loop which is responsible for running the decoder, and an outer stream loop which is responsible for caching the last states of the decoder. During the inference, we start with the outer loop, read the cached states, use them as input to the inner loop and run 5 steps of the decoder, generating 10 Mel-spectrogram frames (100 ms of speech), and finally cache the last states into the state queue. We repeat this process until the inference is complete. As a result, there are always 5 executions of the inner loop per one execution of the outer loop. The sample generation of the WaveRNN starts once it receives the first frame. Additionally, the samples are being played as the Tacotron and WaveRNN are generating them. The streaming architecture greatly reduces the customer perceived latency (CPL) as all the processes are executed in parallel. Fig. 1 depicts the streaming approach in Tacotron during the inference.

### 4.2. Server implementation

On the GPU server, Tacotron uses Tensorflow for the inference engine, whereas WaveRNN uses a persistent CUDA kernel implementation because of its sample-level autoregression. When deploying the end-to-end neural TTS system, we load Tacotron and WaveRNN models on separate GPUs. In addition, since capacity is a major factor for server deployments, we use half precision (FP16) inference for WaveRNN. This enables us to fit 3 Tacotron or 3 WaveRNN models on a single GPU. Overall, the new streaming system achieves 5x faster than real time performance on server and is able to generate 120k audio samples per second.

### 4.3. On-device implementation

On-device TTS provides multiple advantages over the server implementation, such as better response time, off-line or low-bandwidth usage, smaller cloud resource costs, and improved privacy. In this section, we describe the optimizations made specifically for the on-device implementation.

Latest mobile devices typically have multiple hardware backends, such as a CPU, GPU, and dedicated hardware accelerators, and it is important to dispatch network layers to appropriate backends for the best performance. Our internal on-device deep learning framework supports multiple hardware backends, including CPU, GPU, the neural engine—an 8-core dedicated hardware for energy and power efficient acceleration of neural networks, as well as the CPU accelerators—hardware designed to accelerate machine learning applications. We utilize these different backends during runtime while also leveraging the internal on-device deep learning framework to achieve best possible performance, allowing the neural TTS models to run seamlessly on the new generation of mobile devices.

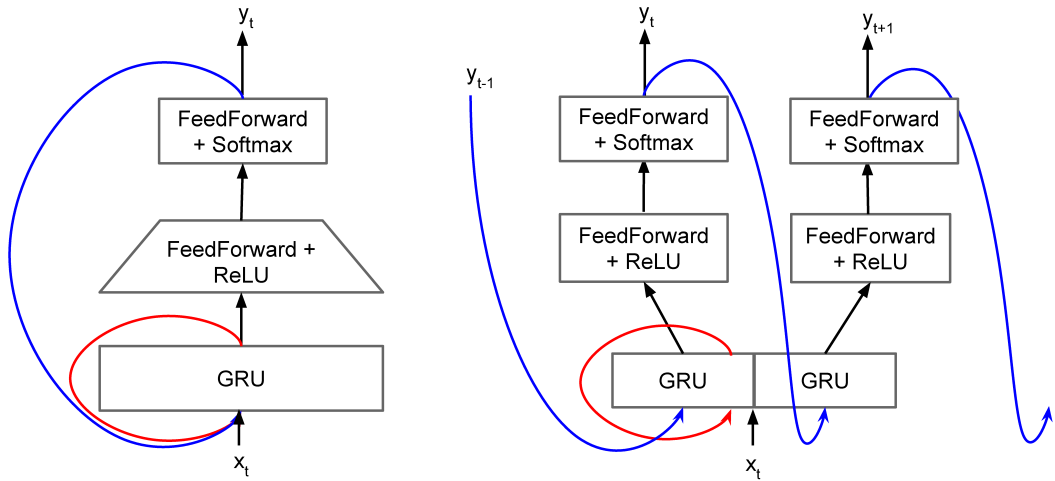

**Fig. 2**. Left: WaveRNN designed for server synthesis. Right: Split-state WaveRNN designed for on-device synthesis. We use only a single GRU where the hidden state is split into two, rather than using two separate GRUs.

*4.3.1. Split-state WaveRNN*

To increase the speed of generation on the device, we propose to use a split-state WaveRNN, where we split the hidden state of the GRU in a way that the first half predicts the current sample and the second half predicts the next sample, as depicted in Fig. 2. Computationally, the most expensive operation in WaveRNN is the GRU hidden state computation which involves the multiplication of a matrix of 1536×512 and a hidden state vector of 512 dimensions. If we store the linear hidden state of the multiplication prior to applying the non-linearity, we can re-use the value during computation of the hidden state for the second sample. This simple method, inspired by the coarse-fine division of the original WaveRNN [3] on the hidden states, reduces the computation by half, or effectively reduces the WaveRNN size to 256 for a single sample prediction. Although the above method may not be as useful on parallel architectures such as GPUs due to the sequential dependency between the sample generation (synchronization delays are the main bottleneck on GPU rather than the compute), it is useful on sequential compute hardware such as CPU.

*4.3.2. WaveRNN in neural engine*

The latest generation of mobile devices typically have a dedicated neural network hardware (neural engine) which allows energy and power efficient inference compared to the CPU. We therefore explored the slow and power consuming inference of WaveRNN (due to sample-level autoregression) on a neural engine. The network has to be compiled to a hardware representation before it can be executed. By utilizing pre-compilation of the layers and caching the dispatch information, we significantly reduce the load time of WaveRNN, and allow the network to be executed with the least amount of overhead on the neural engine, thus reducing the latency.

In the neural engine, we unrolled the WaveRNN sample loop by stacking up a single iteration of the loop 120 times to synthesize one frame of 240 samples. Unrolling avoids making multiple round trips to the driver and back, essentially streamlining the processing on the neural engine for each frame, giving us a significant performance and power usage improvement. The unrolled WaveRNN with 6726 layers is the deepest neural network ever run on this specific neural engine.

The split-state WaveRNN architecture uses a combination of softmax and multinomial sampling layers to generate the final samples. However, this neural engine does not support random number generation and softmax operation. To overcome this, we use the Gumbel-Max trick to sample from the softmax [24, 25]. In the Gumbel-Max trick, the softmax output probabilities are simply added to a pre-sampled Gumbel distribution vector and then argmax is used to sample from the resulting distribution. This eliminates the need to compute cumulative distribution function which is not supported on the neural engine, while the argmax is supported. Since the random sampling layer at the output is independent of other network operations, it is possible to use parallel generation for the random numbers on the CPU while the network is in-flight on the neural engine.

*4.3.3. Optimizations for CPU accelerator*

We also explored the inference of Tacotron and WaveRNN on the CPU accelerators of mobile devices. A CPU accelerator is specially designed to accelerate machine learning applications. The three main optimizations that we applied to run neural TTS models on a CPU accelerator are caching, mixed-precision inference, and resource allocation. The Tacotron output is a sequence of Mel-spectrogram frames, and each frame is processed by WaveRNN to synthesize 240 samples. In order to synthesize 240 samples, the split-state GRU is executed 120 times, generating 2 samples in each step. Since the Mel-spectrogram conditioning remains constant while generating samples within a frame, we cached the output of the first fully connected layer of the WaveRNN to improve performance by 15%.

CPU accelerators typically have high-performance cores (p-cores) and high-efficiency cores (e-cores). Tacotron outputs are at the frame-level, so it is faster than the WaveRNN, which is a sample-level autoregressive network. To optimally utilize both resources, we dispatch Tacotron on e-cores and WaveRNN on p-cores, which improves our real time factor (RTF) to 3x on the CPU.

To further improve the usage of memory and computation on CPU, we utilized mixed precision inference for Tacotron and WaveRNN. More precisely, we use 16-bit floating point accumulation for the fully connected layers and mixed-precision inference for the recurrent and non-linear layers. This improves the performance of the fully connected layers by 2x compared to 32-bit floating point computations. Finally, to match the hardware, we use column-major format which further improves the performance of Tacotron from 5.4x to 8.5x faster than real time.

We also prioritize CPU scheduling for neural TTS to make sure there is no interruption during real-time synthesis when the device is under heavy workload. The results in Section 5.2 show that the performance of the models on CPU accelerators are significantly better than on the neural engine.

## 5. EXPERIMENTS AND RESULTS

### 5.1. Data

We use two internal datasets for our experiments, an American English female dataset (en-US Female) and an American English male dataset (en-US Male). The American English female dataset contains 36-hours of audio, whereas the male dataset contains 22-hours of audio. The speech data is sampled at 24 kHz and the Mel-spectrograms are computed from the pre-emphasized speech using short-time Fourier transform (STFT) with a 25 ms frame length and a 10 ms frame shift. We use 80 Mel filterbanks. The input to Tacotron is phonemes with word boundaries and punctuations. Tacotron and WaveRNN are trained separately and combined during inference.

**Table 1**. Performance results for two mobile devices.

|  | Mobile device A | Mobile device B |
|---|---|---|
| CPL (ms) | 179 | 158 |
| End-to-end RTF | 1.47x | 3.03x |
| Tacotron RTF | 5.5x | 8.5x |
| WaveRNN RTF | 1.6x | 3.3x |

**Table 2**. Disk and memory footprint for hybrid unit-selection, Tacotron, and WaveRNN models.

|  | Unit-sel. | Tacotron | WaveRNN |
|---|---|---|---|
| Disk footprint | 458 MB | 70 MB | 2.5 MB |
| Memory footprint | 2.6 MB | 7.8 MB | 3.2 MB |

### 5.2. Performance

Low latency is critical for making a TTS system responsive to the user. To measure the performance, we use customer perceived latency (CPL), defined as the difference between the time of the TTS request and the time when the first audio buffer is played (lower CPL is better). We also define real-time factor (RTF) as the ratio of total audio duration and total end-to-end synthesis latency (higher RTF is better). A high RTF creates more room for streaming in case of intensive device usage.

The performance results for two different mobile devices are shown in Table 1, where mobile device A refers to the case where Tacotron and WaveRNN inference is performed on the CPU and neural engine, respectively, and B refers to the case where Tacotron and WaveRNN inference is performed on the CPU accelerator. The results show that both mobile devices achieve a CPL of less than 180 ms and a RTF greater than 1.5x. Furthermore, with inference on the CPU accelerator we improve the speed of both models and achieve RTF of 3.3x.

We also report the disk and memory footprints of the hybrid unit-selection system, Tacotron, and WaveRNN in Table 2. The unit-selection system has the largest disk footprint as it has to store the speech units in the database, while the neural systems mostly need to store the network weights. This significantly reduces the required effort to deploy new or updated TTS models to millions of mobile devices. The combined memory usage of Tacotron and WaveRNN is higher than in the unit-selection system, but not prohibitive for mobile devices.

### 5.3. Quality

We compare the neural TTS system with the hybrid unit-selection system trained on the same datasets in order to demonstrate the benefit of using a neural TTS system. The modeling and performance improvements to the neural TTS system introduced in this work were made at no or little cost to quality. For each system, a 5-point mean opinion score (MOS) test was performed on 200 sentences by American English native speakers. Each sentence was rated by at least 15 listeners. Table 3 presents the MOS test results. Results show that the neural TTS system achieves MOS values significantly higher than the hybrid unit-selection system, and the quality gap between the natural and synthetic speech is greatly reduced.

**Table 3**. MOS test results from hybrid unit-selection, neural TTS, and natural speech with 95% confidence intervals.

|                      | en-US Female    | en-US Male      |
|----------------------|-----------------|-----------------|
| Hybrid unit-selection | 3.70 ± 0.037   | 3.87 ± 0.034   |
| Neural TTS           | 4.37 ± 0.026    | 4.35 ± 0.026    |
| Natural              | 4.63 ± 0.032    | 4.62 ± 0.032    |

**Table 4**. MOS test results from hybrid unit-selection and neural TTS for in-domain and out-of-domain test sets.

|        |                | in-domain | out-of-domain |
|--------|----------------|-----------|---------------|
| Female | Unit-selection | 4.15      | 3.44          |
| Female | Neural TTS     | 4.53      | 4.28          |
| Male   | Unit-selection | 4.1       | 3.70          |
| Male   | Neural TT      | 4.48      | 4.28          |

Our test set of 200 sentences includes both in-domain and out-of-domain sentences. In-domain sentences are similar to our training data whereas out-of-domain sentences include sentences from the web, such as Wikipedia and news articles. Table 4 shows that, while the neural TTS system achieves higher MOS than the hybrid unit-selection system for both in-domain and out-of-domain sentences, the improvement is larger for the out-of-domain sentences. The results show the robustness of the neural TTS system for the input text.

## 6. CONCLUSIONS

In this paper, we presented an end-to-end speech synthesis system that is based on Tacotron and WaveRNN models. To our knowledge, it is the first of its kind that runs on new generations of mobile devices. The high quality of the generated speech and the 3x faster than real time generation speed allows this system to be launched as a successful product on mobile devices. To reach both high quality and performance, we made various optimizations and engineering designs. We modified the attention mechanism in Tacotron to improve the quality and robustness of output speech. We developed a streaming technique that was not limited to these architectures, and was extended to the whole chain of modules in the final TTS product. We optimized the GRU size, used 8-bit $\mu$-law quantization and a split-state method to improve the performance of WaveRNN. Finally, we performed optimization at every level of the software stack which drew the best performance from the on-device deep learning framework and neural engine of mobile devices. The outcome is an end-to-end speech synthesis system that runs seamlessly on a mobile device 3x faster than real time and generates high quality synthetic speech for a voice assistant. In future work, we aim to optimize parallel TTS architectures for on-device synthesis for further improvements in performance and quality.